\documentclass
[floatfix,superscriptaddress,secnumarabic,amssymb,amsmath,nobibnotes,aps,prd,showkeys,showpacs,nofootinbib,onecolumn,notitlepage,12pt]{revtex4}%
\usepackage{setspace}
\usepackage{color}
\usepackage{amsmath}
\usepackage{amsfonts}
\usepackage{verbatim}
\usepackage{amssymb}
\usepackage{graphicx,bm}
\usepackage{graphicx}
\usepackage{amsmath}
\usepackage{amssymb}
\usepackage{amssymb}
\usepackage{graphicx,bm}
\usepackage{graphicx}
\usepackage[caption=false]{subfig}%
\providecommand{\U}[1]{\protect\rule{.1in}{.1in}}

\newcommand{\be}{\begin{equation}}
\newcommand{\ee}{\end{equation}}

\newcommand{\mincir}{\raise
-3.truept\hbox{\rlap{\hbox{$\sim$}}\raise4.truept\hbox{$<$}\ }}
\newcommand{\magcir}{\raise
-3.truept\hbox{\rlap{\hbox{$\sim$}}\raise4.truept\hbox{$>$}\ }}

\ifx\pdfoutput\relax\let\pdfoutput=\undefined\fi
\newcount\msipdfoutput
\ifx\pdfoutput\undefined\else
\ifcase\pdfoutput\else
\ifx\paperwidth\undefined\else
\ifdim\paperheight=0pt\relax\else\pdfpageheight\paperheight\fi
\ifdim\paperwidth=0pt\relax\else\pdfpagewidth\paperwidth\fi
\fi\fi\fi
\begin{document}
\title{Chameleon Mechanism in Scalar Nonmetricity Cosmology}
\author{Andronikos Paliathanasis}
\email{anpaliat@phys.uoa.gr}
\affiliation{Institute of Systems Science, Durban University of Technology, Durban 4000,
South Africa}
\affiliation{Departamento de Matem\'{a}ticas, Universidad Cat\'{o}lica del Norte, Avda.
Angamos 0610, Casilla 1280 Antofagasta, Chile}

\begin{abstract}
We investigate the dynamics and the phase-space evolution for the scalar
nonmetricity cosmology with a Chameleon mechanism. In particular, we consider
a spatially flat Friedmann--Lema\^{\i}tre--Robertson--Walker geometry and
within the framework of scalar nonmetricity theory, we consider a
generalization of the Brans-Dicke theory in nonmetricity gravity. Introducing
a pressureless gas as the matter source, we also incorporate a coupling
function responsible for the interaction. Our findings reveal that the choice
of connection in nonmetricity gravity significantly impacts the interaction
between the scalar field and the matter source. For one particular connection,
we discover the absence of asymptotic solutions with a nontrivial interacting
component. More precisely, in this scenario, the matter source does not
directly interact with the scalar field; however, there is an interaction with
the dynamical degrees of freedom provided by the connection.

\end{abstract}
\keywords{Symmetric teleparallel; nonmetricity gravity; scalar field; Chameleon
mechanism; Interaction}\maketitle

\section{Introduction}

\label{sec1}

Scalar nonmetricity theory \cite{sc1} is a generalization of Symmetric
Teleparallel Equivalent General Relativity (STEGR) \cite{Nester:1998mp}, where
a scalar field nonminimally coupled to gravity is introduced in the
gravitational Action Integral. This scalar field interacts with the
nonmetricity scalar through a coupling function. The scalar nonmetricity
theory serves as the analogue of scalar tensor theory and scalar torsion
theory in symmetric teleparallel theory gravity, see the discussion in
\cite{trinity}. While STEGR is equivalent to General Relativity, the presence
of a minimally coupled scalar field maintains this equivalence. However, in
the case of a nonminimally coupled scalar field, there is no equivalence
between scalar nonmetricity and scalar tensor theories.

In the $f\left(  R\right)  $ generalization of General Relativity
\cite{Buda,BOtt}, the new geometrical degrees of freedom can be attributed to
a scalar field, and the gravitational theory is equivalent to a Brans-Dicke
theory \cite{Brans} with zero Brans-Dicke parameter, also known as the
O'Hanlon theory \cite{Sotiriou,odin1,ff1}. In this work, we are interested in
the Brans-Dicke extension in nonmetricity theory \cite{anpal01}. This theory
has the property of attributing the geometrical degree of freedom provided by
the $f\left(  Q\right)  $ generalization of STEGR
\cite{Koivisto2,Koivisto3,rev10,fq1,fq2}. While the scalar tensor theories are
mathematically related through conformal transformations, this is not the case
for scalar nonmetricity theories. Conformal transformations in symmetric
teleparallel gravity lead to gravitational theories where a boundary term
coupled to the scalar field contributes to the gravitational dynamics. This
implies that a nonminimally coupled scalar field in nonmetricity gravity is
not conformally equivalent to a minimally coupled scalar field in STEGR
\cite{gg1}. The effects of conformal transformations on the physical
properties in scalar nonmetricity theory have been studied recently in
\cite{pal2}. It was found that there exists a unique correspondence for the
asymptotic solutions under conformal transformations, and exact solutions
describing cosmic acceleration remain invariant under the transformation that
relates these conformally equivalent theories.

Constraints of $f\left(  Q\right)  $-gravity with cosmological observations
reveal that the alternative theory challenge the standard $\Lambda$-cosmology
\cite{fot1}, a similar result was found in \cite{fot2,fot3}. For an extended
analysis we refer the reader to \cite{fot4} where the inertia effects was
considered in the analysis of the observational data. 

Within this study, we postulate an interaction mechanism between the
nonminimally coupled scalar field and the matter component
\cite{int1,int2,int3,int4,int5,int6,int7}. This interaction is because of a
Chameleon component in the gravitational Action Integral \cite{ch1,ch2}.\ The
matter source Lagrangian is coupled to the scalar field through an interacting
functions. Because of this function there is mass transfer between the scalar
field and the matter source. Hence, the mass of the scalar field depends on
the density of the matter source, which means that the mass of the scalar
field is different in open space and in compact objects, which gives a
Chameleon behaviour to the scalar field. The Chameleon mechanism in General
Relativity can have geometric origin and specifically it is related to the
Weyl Integrable Spacetime theory \cite{salim96}. The Chameleon mechanism is a
particular case of the general interaction theory and it has been widely
applied in gravitational physics and cosmology
\cite{va5,va6,hot,hot2,ame1,ame2,pavon}.

The cosmological Chameleon mechanism can be observable in strong gravitational
effects, specifically the cosmological chameleon may affects the orbits in the
solar system \cite{sol2}, see also the discussion in \cite{sol3}. Parameters
of the Chameleon mechanism have been constrained by the analysis of the data
provided by CAST \cite{sol4a} and XENON1T data \cite{sol4b}. There are various
tests which have been proposed in the literature for the investigation of the
chameleon mechanism, we refer the reader to \cite{sol5,sol6} and references
therein. Therefore the study of a Chameleon mechanism in the modified theories
of gravities and specifically in the extension of the STEGR with a scalar
field is of special interest.

In this study we focus on the cosmological framework of a spatially flat
Friedmann--Lema\^{\i}tre--Robertson--Walker (FLRW) geometry, and we
investigate the dynamics and the asymptotic solutions in scalar nonmetricity
theory with a Chameleon component. For the matter source, we consider a
pressureless dust fluid source which represents the dark matter component of
the universe. In STEGR, the definition of the gravitational Lagrangian depends
on the choice of the connection, which differs from the Levi-Civita connection
\cite{rev10}. Specifically, the connection is considered to be flat and
symmetric, leading to a non-unique definition for the connection. In STEGR,
the selection of the connection does not impact the gravitational theory.
However, this is not the case in extensions of STEGR.

We will observe that for the cosmological model under consideration, the
choice of the connection plays a crucial role in the presence of the Chameleon
mechanism in cosmic evolution. Indeed, within the framework of scalar
nonmetricity theory with matter component coupled to the scalar field in the
gravitational Action, it follows that the Chameleon mechanism contributes to
the mass of the scalar field only for specific selection of the connection.
\ This result is different from that provided by the gravitational theories
where the Levi-Civita connection, i.e. extensions of General Relativity
\cite{va5}, or the vierbein fields, i.e. teleparallelism, are applied
\cite{va001}. That is an interesting property of symmetric teleparallel
theory. 

For our study, we employ the method of dynamical system analysis. We utilize
the Hubble normalization approach to determine the existence of asymptotic
solutions in the phase-space. This approach has been extensively studied in
cosmology and has yielded many important results. It has been instrumental in
deriving selection rules for the viability of gravitational models and
understanding the initial condition problem \cite{dn1,dn2,dn3,dn4,dn5}.
Regarding nonmetricity theory, dynamical system analysis has also been widely
applied, as seen in \cite{df1,df2,df3,df4,df5} and references therein. The
structure of the paper is outlined as follows.

In Section \ref{sec2}, we introduce the basic framework of STEGR. The Scalar
nonmetricity theory is presented in Section \ref{sec3}, where we focus on the
Brans-Dicke analogue in nonmetricity theory. The Chameleon mechanism, which
describes the interaction between the scalar field and the matter source, is
introduced in Section \ref{sec4}. The cosmological application of the
Chameleon mechanism in scalar nonmetricity theory is discussed in Section
\ref{sec5}, where we consider the background geometry to be the spatially flat
FLRW spacetime. We consider the three different families $\Gamma_{1}%
,~\Gamma_{2}$ and $\Gamma_{3}$ for the connection given by STEGR, and we
present three different sets of field equations. We find that only for
connections $\Gamma_{1}$ and $\Gamma_{3}$ does the mass of the scalar field
depend on the Chameleon mechanism, while for connection $\Gamma_{2}$, the
Chameleon mechanism introduces an interaction between the matter source and
the geometrical dynamical degrees of the connection. For connection
$\Gamma_{2}$, there is no dynamical interaction between the scalar field and
the matter source.

In Section \ref{sec6}, we conduct a detailed analysis of the phase-space. We
apply the Hubble normalization approach and express the field equations in an
equivalent form of an algebraic-differential system. We determine the
stationary points and investigate the physical properties of the asymptotic
solutions at these points. Furthermore, we discuss the stability properties of
these asymptotic solutions. Specifically, for connection $\Gamma_{2}$, we find
no asymptotic solutions with nontrivial Chameleon mechanism. However, for
connections $\Gamma_{1}$ and $\Gamma_{3}$, we identify stationary points where
the Chameleon mechanism significantly impacts the universe's evolution. These
points can act as sources or attractors, indicating that the Chameleon
mechanism in scalar nonmetricity cosmology can play an important role in both
early-time and late-time stages of the universe. The case of arbitrary
potential is discussed in Section \ref{sec7a}. Finally, in Section \ref{sec7},
we present our conclusions.

\section{Symmetric teleparallel Gravity}

\label{sec2}

We consider a four-dimensional manifold which is described by the metric
tensor $g_{\mu\nu}$ and the symmetric and flat connection $\Gamma_{\mu\nu
}^{\kappa}$. Therefore, the curvature and the torsion tensors defined for the
connection $\Gamma_{\mu\nu}^{\kappa}$ are always zero, that is,
\cite{trinity}
\begin{align}
R_{\;\lambda\mu\nu}^{\kappa}  &  \equiv2\left(  \Gamma_{\;\lambda\left[
\nu,\mu\right]  }^{\kappa}+\Gamma_{\;\lambda\lbrack\nu}^{\sigma}\Gamma
_{\;\mu]\sigma}^{\kappa}\right)  =0,\\
\mathrm{T}_{\mu\nu}^{\lambda}  &  \equiv2\Gamma_{\;\left[  \mu\nu\right]
}^{\lambda}=0.
\end{align}

However, the nonmetricity scalar defined
\begin{equation}
Q_{\lambda\mu\nu}\equiv g_{\mu\nu,\lambda}-\Gamma_{\;\lambda\mu}^{\sigma
}g_{\sigma\nu}-\Gamma_{\;\lambda\nu}^{\sigma}g_{\mu\sigma},
\end{equation}
plays an important role in nonmetricity gravity.

We define the non-metricity conjugate tensor \cite{Koivisto2}
\begin{equation}
P_{\;\mu\nu}^{\lambda}=-\frac{1}{4}Q_{\;\mu\nu}^{\lambda}+\frac{1}{2}%
Q_{(\mu\phantom{\lambda}\nu)}^{\phantom{(\mu}\lambda\phantom{\nu)}}+\frac
{1}{4}\left(  Q^{\lambda}-\bar{Q}^{\lambda}\right)  g_{\mu\nu}-\frac{1}%
{4}\delta_{\;(\mu}^{\lambda}Q_{\nu)} \label{defP}%
\end{equation}
in which $Q_{\lambda}=Q_{\lambda\mu}^{\phantom{\lambda\mu}\mu}$, $\bar
{Q}_{\lambda}=Q_{\phantom{\mu}\lambda\mu}^{\mu}$. The nonmetricity
$Q=Q_{\lambda\mu\nu}P^{\lambda\mu\nu}~$is the fundamental scalar of STEGR. The
gravitational Action in STEGR is defined as%
\begin{equation}
S_{Q}=\int d^{4}x\sqrt{-g}Q\,.
\end{equation}
However, the nonmetricity scalar $Q$ is related to the Ricci scalar $\tilde
{R}$ defined by the Levi-Civita connection $\tilde{\Gamma}_{\mu\nu}^{\kappa}$
of the metric tensor, with the relation~$\tilde{R}-Q=B_{Q},$where $B_{Q}$ is
defined by the expression \cite{gg1} $B_{Q}=-\tilde{\nabla}_{\mu}\left(
Q^{\mu}-\bar{Q}^{\mu}\right)  $\thinspace\ and it is a boundary term.
Therefore, the Action Integral for STEGR reads%
\begin{equation}
\int d^{4}x\sqrt{-g}Q\simeq\int d^{4}x\sqrt{-g}\tilde{R}+\text{boundary
terms.}%
\end{equation}

\section{Scalar nonmetricity Gravity}

\label{sec3}

In scalar nonmetricity theory we assume the existence of a scalar field $\phi
$, coupled to the gravity where the gravitational Action Integral reads
\cite{sc1}
\begin{equation}
S_{ST\varphi}=\int d^{4}x\sqrt{-g}\left(  \frac{F\left(  \varphi\right)  }%
{2}Q+\frac{\omega\left(  \varphi\right)  }{2}g^{\mu\nu}\varphi_{,\mu}%
\varphi_{,\nu}+V\left(  \varphi\right)  \right)  , \label{ac.01}%
\end{equation}
$F\left(  \varphi\right)  $ is the coupling function of the scalar field with
the nonmetricity scalar, and $V\left(  \varphi\right)  $ is the scalar field
potential. Function $\omega\left(  \phi\right)  $ defines only the
\textquotedblleft coordinate\textquotedblright\ where the scalar field is
defined, and it can be removed from the system with the introduction of the
new scalar field $d\Phi=\sqrt{\omega\left(  \varphi\right)  }d\varphi
~$\cite{anpal01}.

Variation with respect to the metric tensor in (\ref{ac.01}) gives the
modified field equations \cite{jjd1}%

\begin{equation}
F\left(  \varphi\right)  G_{\mu\nu}+2F_{,\phi}\varphi_{,\lambda}P_{~~\mu\nu
}^{\lambda}+g_{\mu\nu}V\left(  \varphi\right)  +\frac{\omega\left(
\varphi\right)  }{2}\left(  g_{\mu\nu}g^{\lambda\kappa}\varphi_{,\lambda
}\varphi_{,\kappa}-\varphi_{,\mu}\varphi_{,\nu}\right)  =0. \label{ai.001}%
\end{equation}
where $G_{\mu\nu}$ is the Einstein tensor of general relativity. Moreover, the
Klein-Gordon equation for the scalar field reads%
\begin{equation}
\frac{\omega\left(  \varphi\right)  }{\sqrt{-g}}g^{\mu\nu}\partial_{\mu
}\left(  \sqrt{-g}\partial_{\nu}\varphi\right)  +\frac{\omega_{,\varphi}}%
{2}g^{\lambda\kappa}\varphi_{,\lambda}\varphi_{,\kappa}+\frac{1}{2}%
F_{,\varphi}Q-V_{,\varphi}=0. \label{ai.003}%
\end{equation}
Finally, variation of (\ref{ac.01}) with respect to the connection$~\Gamma
_{\mu\nu}^{\kappa}$ gives the equation of motion
\begin{equation}
\nabla_{\mu}\nabla_{\nu}\left(  \sqrt{-g}F\left(  \varphi\right)
P_{\phantom{\mu\nu}\sigma}^{\mu\nu}\right)  =0. \label{ai.002}%
\end{equation}
When the latter equation is trivially satisfied, we shall say that the
connection $\Gamma_{\mu\nu}^{\kappa}$ is defined in the coincidence gauge.

In the following we consider the $F\left(  \varphi\right)  =\varphi$ and
$\omega\left(  \varphi\right)  =\frac{\omega_{0}}{\varphi}$, where the Action
(\ref{ac.01})
\begin{equation}
S_{BD\varphi}=\int d^{4}x\sqrt{-g}\left(  \frac{\varphi}{2}Q+\frac{\omega
}{2\varphi}g^{\mu\nu}\varphi_{,\mu}\varphi_{,\nu}+V\left(  \varphi\right)
\right)  , \label{ai.002a}%
\end{equation}
and takes the form of the Brans-Dicke analogue in nonmetricity theory. When
parameter $\omega=0$, the gravitational theory (\ref{ai.002a}) is equivalent
to the symmetry teleparallel $f\left(  Q\right)  $-gravity, where the scalar
field $\varphi$ attributes the new dynamical degrees of freedom provided by
the nonlinear function $f\left(  Q\right)  $. Specifically, the scalar field
is introduced as $\varphi=f_{,Q}\left(  Q\right)  $ and $V\left(
\varphi\right)  =f_{,Q}\left(  Q\right)  Q-f\left(  Q\right)  $.

We employ the change of variable $\varphi=e^{\phi}$ and the Brans-Dicke
(-like) Action Integral (\ref{ai.002}) takes the form of the dilaton analogue
in nonmetricity theory, that is, \cite{anpal01}
\begin{equation}
S_{D}=\int d^{4}x\sqrt{-g}e^{\phi}\left(  \frac{Q}{2}+\frac{\omega}{2}%
g^{\mu\nu}\phi_{,\mu}\phi_{,\nu}+\hat{V}\left(  \phi\right)  \right)
~\text{\ with }\hat{V}\left(  \phi\right)  =V\left(  \phi\right)  e^{-\phi}.
\label{dil.01}%
\end{equation}
for the dilaton field the gravitational field equations are
\begin{equation}
G_{\mu\nu}+2\phi_{,\lambda}P_{~~\mu\nu}^{\lambda}+g_{\mu\nu}V\left(
\phi\right)  e^{-\phi}+\frac{\omega}{2}\left(  g_{\mu\nu}g^{\lambda\kappa}%
\phi_{,\lambda}\phi_{,\kappa}-\phi_{,\mu}\phi_{,\nu}\right)  =0.
\end{equation}%
\begin{equation}
\omega\nabla_{\mu}\nabla^{\mu}\phi+\frac{1}{2}Q-V_{,\phi}e^{-\phi}=0.
\end{equation}%
\begin{equation}
\nabla_{\mu}\nabla_{\nu}\left(  \sqrt{-g}e^{\phi}P_{\phantom{\mu\nu}\sigma
}^{\mu\nu}\right)  =0.
\end{equation}

Conformal transformations of the cosmological nonmetricity Brans-Dicke theory
have been investigated in \cite{pal2}. It was found that there exists a unique
one-to-one correspondence for asymptotic solutions between different frames,
the correspondence is independent on the selection of the connection. That
result differs from that of metric Brans-Dicke theory, which means that
nonmetricity theory opens new directions on the discussion of the physical frame.

The nonmetricity scalar $Q$ is related to the Ricci scalar $R$ for the
Levi-Civita connection of the metric tensor $g_{\mu\nu}$, with a boundary
term, that is $Q=R+B~$\cite{fbb}. Consequently, by replacing in (\ref{ac.01})
it follows that the theory deviates from the usual scalar-tensor model. 

\section{Chameleon Mechanism}

\label{sec4}

Let us now introduce a matter component in the gravitational Action Integral
for the dilaton field (\ref{dil.01}) such that
\begin{equation}
S=S_{D}+\int\sqrt{-g}g\left(  \phi\right)  L_{m}%
\end{equation}
where $g\left(  \phi\right)  $ is the coupling function between the matter
source and the scalar field $\phi$. A nonconstant function $g\left(
\phi\right)  $ indicates that there exists energy transfer between the matter
source and the scalar field $\phi$. Lagrangian $L_{m}$ describes the matter term.

The gravitational field equations read%
\begin{equation}
G_{\mu\nu}=g\left(  \phi\right)  T_{\mu\nu}^{m}+T_{\mu\nu}^{\phi}~,~
\label{mfe.01}%
\end{equation}%
\begin{equation}
\nabla_{\mu}\nabla_{\nu}\left(  \sqrt{-g}e^{\phi}P_{\phantom{\mu\nu}\sigma
}^{\mu\nu}\right)  =0.
\end{equation}
where\ tensor $T_{\mu\nu}^{\phi}$ attributes the scalar field dynamical terms
and $T_{\mu\nu}^{m}$ is the energy momentum tensor for the matter source. For
the comoving observer~$u_{\mu}$ with $g_{\mu\nu}u^{\mu}u^{\nu}=-1$, the energy
momentum tensor for a perfect fluid is
\begin{equation}
T_{\mu\nu}^{m}=\left(  \rho_{m}+p_{m}\right)  u_{\mu}u_{\nu}+p_{m}g_{\mu\nu},
\end{equation}
where $\rho_{m}$ is the energy density and $p_{m}$ is the pressure component.

The equation of motion for the matter source reads%
\begin{equation}
\tilde{\nabla}_{\nu}\left(  g\left(  \phi\right)  T^{m~\mu\nu}+T^{\phi~\mu\nu
}\right)  =0
\end{equation}
where $\tilde{\nabla}_{\nu}$ is defined by the Levi-Civita connection. We
remark that we have assumed no interaction of the matter source with the
connection $\Gamma_{\mu\nu}^{\kappa}$. The latter equation of motion reads%
\begin{equation}
\tilde{\nabla}_{\nu}\left(  f\left(  \phi\right)  T^{m~\mu\nu}\right)
=f_{,\phi}\rho_{m}~,~\tilde{\nabla}_{\nu}\left(  T^{\phi~\mu\nu}\right)
=-f_{,\phi}.
\end{equation}

Therefore, for the perfect fluid and the scalar field we find the modified
equations of motions%
\begin{equation}
\tilde{\nabla}_{\mu}\left(  \rho_{m}\right)  u^{\mu}+\left(  \rho_{m}%
+p_{m}\right)  \tilde{\nabla}_{\mu}u^{\mu}-\rho_{m}\tilde{\nabla}_{\mu}%
\ln\left(  g\left(  \phi\right)  \right)  u^{\mu}=0,
\end{equation}%
\begin{equation}
\tilde{\nabla}^{\nu}\left(  2e^{\phi}\phi_{,\lambda}P_{~~\mu\nu}^{\lambda
}+g_{\mu\nu}V\left(  \phi\right)  +\frac{\omega}{2}e^{\phi}\left(  g_{\mu\nu
}g^{\lambda\kappa}\phi_{,\lambda}\phi_{,\kappa}-\phi_{,\mu}\phi_{,\nu}\right)
\right)  +e^{\phi}\phi^{,\nu}G_{\mu\nu}+2\rho_{m}\tilde{\nabla}_{\mu}g\left(
\phi\right)  =0.
\end{equation}

In the following we consider the exponential coupling function
\begin{equation}
g\left(  \phi\right)  =g_{0}e^{g\phi}. \label{cp1}%
\end{equation}
For this model, the latter equations of motion are simplified as follows%
\begin{equation}
\tilde{\nabla}_{\mu}\left(  \rho_{m}\right)  u^{\mu}+\left(  \rho_{m}%
+p_{m}\right)  \tilde{\nabla}_{\mu}u^{\mu}-\lambda\rho_{m}\phi_{,\mu}u^{\mu
}=0, \label{ds1}%
\end{equation}%
\begin{equation}
\tilde{\nabla}^{\nu}\left(  2e^{\phi}\phi_{,\lambda}P_{~~\mu\nu}^{\lambda
}+g_{\mu\nu}V\left(  \phi\right)  +\frac{\omega}{2}e^{\phi}\left(  g_{\mu\nu
}g^{\lambda\kappa}\phi_{,\lambda}\phi_{,\kappa}-\phi_{,\mu}\phi_{,\nu}\right)
\right)  +e^{\phi}\phi^{,\nu}G_{\mu\nu}+2\rho_{m}\lambda e^{\lambda\phi}%
\phi_{,\mu}=0. \label{ds2}%
\end{equation}
The exponential coupling function describes the simplest interacting scenario
and it is the toy model which has been applied before in the case of General
Relativity \cite{ch1,ch2}.

\section{Interaction in Nonmetricity Cosmology}

\label{sec5}

In large scales the universe is considered to be homogeneous and isotropic
described by the spatially flat FLRW spacetime with line element
\begin{equation}
ds^{2}=-N(t)^{2}dt^{2}+a(t)^{2}\left[  dr^{2}+r^{2}\left(  d\theta^{2}%
+\sin^{2}\theta d\varphi^{2}\right)  \right]  , \label{genlineel}%
\end{equation}
where $a\left(  t\right)  $ is the scale factor denotes the radius of the
universe and $N\left(  t\right)  $ is the lapse function. The Hubble function
is defined as $H=\frac{1}{N}\frac{\dot{a}}{a}$, where dot means total
derivative with respect to the $t~$parameter, i.e.~$\dot{a}=\frac{da}{at}$.
The FLRW geometry for arbitrary scale factor $a\left(  t\right)  $ admits a
sixth-dimensional Killing algebra consisted by the six isometries of the
three-dimensional Euclidean space, three translations and three rotations.

We introduce a symmetric connection $\Gamma_{\mu\nu}^{\kappa}$, and the
requirements the connection to admit the symmetries of the FLRW spacetime and
the curvature tensor to be zero, leads to the following three families of
connections with nonzero coefficients \cite{Hohmann}%
\[
\Gamma_{1}:\Gamma_{\;tt}^{t}=\gamma(t),
\]%
\[
\Gamma_{2}:\Gamma_{\;tt}^{t}=\frac{\ddot{\psi}(t)}{\dot{\psi}(t)}+\dot{\psi
}(t),\quad\Gamma_{\;tr}^{r}=\Gamma_{\;rt}^{r}=\Gamma_{\;t\theta}^{\theta
}=\Gamma_{\;\theta t}^{\theta}=\Gamma_{\;t\varphi}^{\varphi}=\Gamma_{\;\varphi
t}^{\varphi}=\dot{\psi}(t),
\]
and%
\[
\Gamma_{3}:\Gamma_{\;tt}^{t}=\ln\left(  \dot{\Psi}\left(  t\right)  \right)
^{\cdot},\quad\Gamma_{\;rr}^{t}=\frac{1}{\dot{\Psi}\left(  t\right)  }%
,\quad\Gamma_{\;\theta\theta}^{t}=\frac{r^{2}}{\dot{\Psi}\left(  t\right)
},\quad\Gamma_{\;\varphi\varphi}^{t}=\frac{r^{2}\sin^{2}\theta,}{\dot{\Psi
}\left(  t\right)  }%
\]
and common coefficients $\Gamma_{\theta\theta}^{r}=-r~,~\Gamma_{\varphi
\varphi}^{r}=-r\sin^{2}\theta$, $\Gamma_{\varphi\varphi}^{\theta}=-\sin
\theta\cos\theta~,~\Gamma_{\theta\varphi}^{\varphi}=\Gamma_{\varphi\theta
}^{\varphi}=\cot\theta$, $\Gamma_{\;r\theta}^{\theta}=\Gamma_{\;\theta
r}^{\theta}=\Gamma_{\;r\varphi}^{\varphi}=\Gamma_{\;\varphi r}^{\varphi}%
=\frac{1}{r}$. Functions $\gamma\left(  t\right)  ,~\psi\left(  t\right)  $
and $\Psi\left(  t\right)  ~$are dynamical variables \cite{Hohmann}.

Since the three connections are flat, it means that there exist coordinate
transformations in order to be written as $\Gamma_{\mu\nu}^{\kappa}=0$.
However, when we introduce the metric tensor (\ref{genlineel}) we make the
selection for a coordinate system, such that connection $\Gamma_{1}$ to be
defined in the coincidence gauge while connections $\Gamma_{2}$ and
$\Gamma_{3}$ to be defined in the noncoincidence gauge. Consequently, new
dynamical degrees of freedom are introduced in the gravitational field
equations from these two connections.

For the matter source we assume an ideal gas with constant equation of state
parameter, i.e. $p_{m}=w_{m}\rho_{m}$, then for the exponential coupling
function the equation of motion for the matter source (\ref{ds1}) reads%
\begin{equation}
\dot{\rho}_{m}+3\left(  1+w_{m}\right)  H\rho_{m}-\lambda\dot{\phi}\rho_{m}=0
\label{fe.00}%
\end{equation}
from where it follows $\rho_{m}\left(  t\right)  =\rho_{m0}a^{-3\left(
1+w_{m}\right)  }e^{\lambda\phi}$. In the following we consider that the
matter source describes the pressureless dark matter, that is, $w_{m}=0$.

As far as the modified gravitational field equations (\ref{mfe.01}) is
concerned, in the following lines we present the field equations for the
dilaton field (\ref{dil.01}) for the three different connections which
describe the spatially flat FLRW geometry.

Connection $\Gamma_{1}$~is defined in the coincidence gauge and the
corresponding field equations are%

\begin{align}
e^{\phi}\left(  3H^{2}+\frac{\omega}{2}\dot{\phi}^{2}+V\left(  \phi\right)
e^{-\phi}\right)  -g_{0}e^{\lambda\phi}\rho_{m}  &  =0,\label{cn.01}\\
e^{\phi}\left(  2\dot{H}+3H^{2}+2H\dot{\phi}-\frac{\omega}{2}\dot{\phi}%
^{2}+V\left(  \phi\right)  e^{-\phi}\right)   &  =0. \label{cn.02}%
\end{align}
while the Klein-Gordon equation (\ref{ds2}) is expressed as
\begin{equation}
e^{\phi}\left(  2\omega\left(  \ddot{\phi}+3H\dot{\phi}\right)  -6H^{2}%
+\omega\dot{\phi}^{2}+e^{-\phi}V_{,\phi}\right)  +4g_{0}\lambda e^{\lambda
\phi}\rho_{m}=0. \label{cn.03}%
\end{equation}

The field equations for the latter system is equivalent with that of the
scalar-torsion model in teleparallelism with a Chameleon mechanism
\cite{va001}.

For connection $\Gamma_{2}$, the modified Friedmann's equations read%
\begin{align}
e^{\phi}\left(  3H^{2}+\frac{\omega}{2}\dot{\phi}^{2}+\frac{3}{2}\dot{\phi
}\dot{\psi}+V\left(  \phi\right)  e^{-\phi}\right)  -g_{0}e^{\lambda\phi}%
\rho_{m} &  =0,\label{cn.04}\\
e^{\phi}\left(  2\dot{H}+3H^{2}+2H\dot{\phi}-\frac{\omega}{2}\dot{\phi}%
^{2}-\frac{3}{2}\dot{\phi}\dot{\psi}+V\left(  \phi\right)  e^{-\phi}\right)
&  =0,\label{cn.05}%
\end{align}
while the equation of motion for the connection and the scalar field are
\begin{align}
e^{\phi}\left(  3\ddot{\psi}+2\omega\left(  \ddot{\phi}+3H\dot{\phi}\right)
+3\left(  \ddot{\psi}+3H\dot{\psi}\right)  -6H^{2}+\omega\dot{\phi}%
^{2}+2e^{-\phi}V_{,\phi}\right)  +4g_{0}\lambda e^{\lambda\phi}\rho_{m} &
=0,\label{cn.06}\\
\ddot{\phi}+\dot{\phi}^{2}+3H\dot{\phi} &  =0.\label{cn.07}%
\end{align}
From equation (\ref{cn.07}) it follows that $e^{\phi}=\int\frac{\phi_{0}%
}{a^{3}}dt$, which means that the mass of the scalar field does not depend on
the Chameleon mechanism and there is not interaction between the scalar field
and the fluid source. However, equation (\ref{cn.06}) indicates that the
scalar field~$\psi$ which is provided by the connection in the dynamical
system interacts with the Chameleon mechanism.

Finally, for the third connection, $\Gamma_{3}$, the system of the
cosmological field equations is expressed as follows%
\begin{align}
\left(  3H^{2}+\frac{\omega}{2}\dot{\phi}^{2}-\frac{3}{2a^{2}}\frac{\dot{\phi
}}{\dot{\Psi}}+V\left(  \phi\right)  \right)  -g_{0}e^{\lambda\phi}\rho_{m}
&  =0,\label{cn.08}\\
e^{\phi}\left(  2\dot{H}+3H^{2}+2H\dot{\phi}-\frac{\omega}{2}\dot{\phi}%
^{2}-\frac{1}{2a^{2}}\frac{\dot{\phi}}{\dot{\Psi}}+V_{,\phi}e^{-\phi}\right)
&  =0. \label{cn.09}%
\end{align}
The equations of motion for the scalar field and the connection are
\begin{align}
e^{\phi}\left(  -\frac{3}{a^{2}}\left(  \ddot{\Psi}-H\dot{\Psi}\right)
+\dot{\Psi}^{2}\left(  -6H^{2}+\omega\dot{\phi}^{2}+2\omega\left(  \ddot{\phi
}+3H\dot{\phi}\right)  \right)  +2V_{,\phi}e^{-\phi}\dot{\Psi}^{2}\right)
+4g_{0}\lambda e^{\lambda\phi}\rho_{m}  &  =0,\label{cn.10}\\
\dot{\Psi}\left(  \dot{\phi}\left(  H+\dot{\phi}\right)  +\ddot{\phi}\right)
-2\dot{\phi}\ddot{\Psi}  &  =0. \label{cn.11}%
\end{align}

From the modified Klein-Gordon equations (\ref{cn.03}), (\ref{cn.07}) and
(\ref{cn.11}) we observe that the contribution of the Chameleon mechanism in
the mass of the scalar field depends on the selection of the connection for
the definition of the nonmetricity scalar. For the coincidence gauge, and from
the Klein-Gordon equation (\ref{cn.03}) the term provided by the Chameleon
mechanism in the Klein-Gordon equation is similar to that of the metric
theory. Nevertheless, for connection $\Gamma_{2}$, and from the Klein-Gordon
equation (\ref{cn.07}) it follows that the mass of the scalar field is always
zero, while the Chameleon mechanism interacts with the connection. Finally,
for the third connection $\Gamma_{3}$ and equation (\ref{cn.11}) we observe
that the scalar field is given by the expression $e^{\phi}=\int\frac{\phi
_{0}\dot{\Psi}^{2}}{a}d\not t  $, which means that the scalar field depends on
the dynamical variable of the connection; however the scalar field interacts
indirectly with the Chameleon mechanism through the field $\Psi\left(
t\right)  $.

The effects of the Chameleon mechanism in the scalar field mass depends on the
selection of the connection.\ For the coincidence gauge the\ Chameleon
mechanism has similarities with the metric theories, while for the
noncoincidence gauge the Chameleon mechanism interact with the scalar field indirectly.

\section{Phase-space analysis}

\label{sec6}

In this Section we investigate the asymptotic solutions and the general
evolution for the cosmological model of scalar nonmetricity theory with
Chameleon mechanism. For the different cosmological theories related to the
three different connections of nonmetricity theory, we define dimensionless
variables in the Hubble normalization \cite{cope} in order to write the field
equations as a set of algebraic-differential system.

The stationary points of the latter system describe asymptotic cosmological
solutions. From the stability properties of the stationary points we can
construct the global evolution for the dynamical and physical parameters of
the model and to understand the effects of the Chameleon mechanism in the
cosmological evolution. The new set of dependent and independent variables in
the Hubble normalization are
\begin{equation}
x=\frac{\dot{\phi}}{2\sqrt{6}H}~,~y=\frac{e^{-\phi}V\left(  \phi\right)
}{3H^{2}}~,~\Omega_{m}=g_{0}e^{\left(  \lambda-1\right)  \phi}\frac{\rho_{m}%
}{3H^{2}}~,
\end{equation}%
\begin{equation}
\kappa=\frac{V_{,\phi}}{V}~,~v=\sqrt{\frac{3}{2}}\frac{\dot{\psi}}{H}%
~,~\frac{1}{\bar{z}}=\sqrt{\frac{3}{2}}a^{2}\dot{\Psi}H~,~\tau=\ln\left(
a\right)  .
\end{equation}

Let $\tau$ be the new independent variable, then for each connection
$\Gamma_{1},~\Gamma_{2}$ and $\Gamma_{3}$ we can write the corresponding
reduced dynamical system. For the potential function $V\left(  \phi\right)  $
we introduce the exponential function $V\left(  \phi\right)  =V_{0}%
e^{\kappa\phi}$, in such a way that the parameter $\kappa$ to be always a
constant. The analysis for the exponential potential includes the derivation
of the stationary points for every potential function which can be described
in the limit by the exponential potential such that parameter $\kappa$ remains
a constant.

We are interested in the asymptotic limit and the exponential function
provides the asymptotics where the scalar field dominates in the field
equations and also it the case where it does not contribute in the evolution
of the universe. This has been widely discussed before for instance in
\cite{ans11} and references therein.

In this case the dimension of the dynamical system is reduced by one.

For connection $\Gamma_{1}$, we find the dynamical system $DS_{I}$%
\begin{align}
\Omega_{m}  &  =1+\omega x^{2}+y~,\label{ds.01}\\
\frac{dx}{d\tau}  &  =x^{2}\left(  \sqrt{6}\left(  1+2\lambda\right)  -3\omega
x\right)  +3x\left(  y-1\right)  +\frac{\sqrt{6}}{\omega}\left(
1+2\lambda-\left(  \kappa-2\lambda\right)  y\right)  ,\label{ds.02}\\
\frac{dy}{d\tau}  &  =y\left(  \sqrt{6}\left(  1+\kappa\right)  x-3\left(
\omega x^{2}-\left(  1+y\right)  \right)  \right)  . \label{ds.03}%
\end{align}

The second connection $\Gamma_{2}$ provides the dynamical system $DS_{II}$%
\begin{align}
\Omega_{m}  &  =1+\omega x^{2}+y+xz~,\label{ds.04}\\
\frac{dx}{d\tau}  &  =-\frac{3}{2}x\left(  1-y+x\left(  \omega x+z\right)
\right)  ,\label{ds.05}\\
\frac{dy}{d\tau}  &  =y\left(  3\left(  1+y\right)  +x\left(  \sqrt{6}\left(
1+\kappa\right)  -3\left(  \omega x+z\right)  \right)  \right)  ,
\label{ds.06}\\
\frac{dz}{d\tau}  &  =\left(  \sqrt{6}\left(  1+2\lambda\right)  -\frac{3}%
{2}z\right)  \left(  1+\omega x^{2}+xz\right)  +y\left(  \frac{3}{2}z-\sqrt
{6}\left(  \kappa-2\lambda\right)  \right)  . \label{ds.07}%
\end{align}

Finally, for connection $\Gamma_{2}$ the corresponding dynamical system
$DS_{III}~$in the Hubble normalization is determined to be
\begin{align}
\Omega_{m}  &  =1+\omega x^{2}+y+xv,\label{ds.08}\\
\frac{\left(  4\omega x+v\right)  }{x}\frac{dx}{d\tau}  &  =\left(  2\sqrt
{6}\left(  1+2\lambda\right)  +\frac{v}{2}-6\omega x\right)  \left(  1+\omega
x^{2}\right)  +2v~...\nonumber\\
&  ~~~~~...~+y\left(  \frac{3}{2}z-2\sqrt{6}\left(  \kappa-2\lambda\right)
\right)  +x\left(  6\omega y+\left(  4\sqrt{6}\lambda+\frac{v}{2}\right)
v\right)  ,\label{ds.09}\\
\frac{dy}{d\tau}  &  =y\left(  3\left(  1+y\right)  +x\left(  \sqrt{6}\left(
1+\kappa\right)  -3\omega x+v\right)  \right)  ,\label{ds.10}\\
\frac{3\left(  4\omega x+v\right)  }{v}\frac{dv}{d\tau}  &  =2\sqrt{6}\left(
1+2\lambda\right)  +\omega x^{2}\left(  12\omega x-z+2\sqrt{6}\left(
2\lambda-3\right)  \right)  +3z~...\nonumber\\
&  ...~-x\left(  4\omega\left(  1+3y\right)  +z\left(  2\sqrt{6}\left(
1-2\lambda\right)  +z\right)  \right)  -3y\left(  3z+2\sqrt{6}\left(
\kappa-2\lambda\right)  \right)  . \label{ds.11}%
\end{align}

For the above dynamical systems the dynamical variables are not constraint
$\left\{  \Omega_{m},x,y,z,v\right\}  $. For instance the effective energy
density for the matter source $~\Omega_{m}$ can be expressed as $\Omega
_{m}=g_{0}e^{\left(  \lambda-1\right)  \phi}\hat{\Omega}_{m}$, $\hat{\Omega
}_{m}=\frac{\rho_{m}}{3H^{2}}$; nevertheless $0\leq\hat{\Omega}_{m}\leq1$, the
range of the dynamical variable $\Omega_{m}$ depend on the value of $g_{0}$,
and of the scalar $\left(  \lambda-1\right)  \phi$. \ Below we focus in the
case where the dynamical variables take values in the finite regime.

\subsection{Dynamical system $DS_{I}$}

We consider connection $\Gamma_{1}$ where the physical parameters are
described by the dynamical system $DS_{I}\,$. In the following we study the
phase-space of system $DS_{I}$.

The stationary points $A=\left(  x\left(  A\right)  ,y\left(  A\right)
\right)  $ for the dynamical system (\ref{ds.01})-(\ref{ds.03}) describe
asymptotic solutions where the effective cosmological fluid has the equation
of state parameter $w_{eff}^{\Gamma_{1}}\left(  x,y\right)  =\left(
2\sqrt{\frac{2}{3}}-\omega x\right)  x+y$. \ 

Thus, from the dynamical system (\ref{ds.01})-(\ref{ds.03}) we derive the points:

\begin{itemize}
\item
\[
A_{1}=\left(  \sqrt{\frac{2}{3}}\frac{\left(  1+2\lambda\right)  }{\omega
},0\right)  ,
\]
with $\Omega_{m}\left(  A_{1}\right)  =1+\frac{2\left(  1+2\lambda\right)
^{2}}{3\omega}$. The stationary point exist for $\omega\neq0$. It describes a
universe where the scalar field and the matter source contribute in the
cosmological evolution and $w_{eff}^{\Gamma_{1}}\left(  A_{1}\right)
=\frac{2-8\lambda^{2}}{\omega}$.~The physical properties of the asymptotic
solution depend on the coupling parameter $\omega$, and on the parameter
$\lambda$ of the interaction term, thus this asymptotic solution can describe
the cosmic acceleration for specific values of the free parameters. The
eigenvalues of the linearized system (\ref{ds.02}), (\ref{ds.03}) around the
stationary point are $e_{1}\left(  A_{1}\right)  =-\frac{2+8\lambda\left(
1+\lambda\right)  +3\omega}{\omega}$,~$e_{2}\left(  A_{1}\right)
=3+\frac{2\left(  \kappa-2\lambda\right)  \left(  1+2\lambda\right)  }{\omega
}$, from where we conclude that $A_{1}$ is an attractor when $\kappa
<\frac{4\lambda\left(  1+2\lambda\right)  -3\omega}{2\left(  1+2\lambda
\right)  }$ with$~\left\{  \lambda<-\frac{1}{2},\omega<-\frac{2}{3}\left(
1+2\lambda\right)  ^{2}\right\}  $, $\left\{  \lambda>-\frac{1}{2}%
,\omega>0\right\}  $ or $\kappa>\frac{4\lambda\left(  1+2\lambda\right)
-3\omega}{2\left(  1+2\lambda\right)  }$ with $\left\{  \lambda<-\frac{1}%
{2},\omega>0\right\}  $ or $\left\{  \lambda>-\frac{1}{2},\omega<-\frac{2}%
{3}\left(  1+2\lambda\right)  ^{2}\right\}  $.

\item
\[
A_{2}=\left(  \frac{1+\kappa}{\sqrt{6}\omega},-1-\frac{\left(  1+\kappa
\right)  ^{2}}{6\omega}\right)  ,
\]
with $\Omega_{m}\left(  A_{2}\right)  =0$,~$w_{eff}^{\Gamma_{1}}\left(
A_{2}\right)  =-1+\frac{1-\kappa^{2}}{3\omega}$ and eigenvalues $e_{1}\left(
A_{2}\right)  =-3-\frac{\left(  1+\kappa\right)  \left(  \kappa-2\lambda
\right)  }{\omega}$,~$e_{2}\left(  A_{2}\right)  =-3-\frac{\left(
1+\kappa\right)  ^{2}}{2\omega}$. In this asymptotic solution only the scalar
field contributes in the universe, and acceleration is occurred for specific
values of the free parameters $\omega$ and $\kappa$. As far as the stability
of the point is concerned, for $\kappa=-1$, the point is always an attractor,
for $\kappa<-1$ the point is an attractor when $\left\{  \omega<-\frac{\left(
1+\kappa\right)  ^{2}}{6},\lambda<\frac{\kappa\left(  1+\kappa\right)
+3\omega}{2\left(  1+\kappa\right)  }\right\}  $ or $\left\{  \omega
>0,\lambda>\frac{\kappa\left(  1+\kappa\right)  +3\omega}{2\left(
1+\kappa\right)  }\right\}  $ while for $\kappa>-1$, the point is an attractor
for $\left\{  \omega<-\frac{\left(  1+\kappa\right)  ^{2}}{6},\lambda
>\frac{\kappa\left(  1+\kappa\right)  +3\omega}{2\left(  1+\kappa\right)
}\right\}  $ or $\left\{  \omega>0,\lambda<\frac{\kappa\left(  1+\kappa
\right)  +3\omega}{2\left(  1+\kappa\right)  }\right\}  $.

\item
\[
A_{3}=\left(  -\sqrt{\frac{3}{2}}\frac{1}{\kappa-2\lambda},\frac{2\left(
\kappa-2\lambda\right)  \left(  1+2\lambda\right)  +3\omega}{2\left(
\kappa-2\lambda\right)  ^{2}}\right)  ,
\]
with $\Omega_{m}\left(  A_{3}\right)  =\frac{\left(  1+\kappa\right)  \left(
\kappa-2\lambda\right)  +3\omega}{\left(  \kappa-2\lambda\right)  ^{2}}%
$,~$w_{eff}^{\Gamma1}\left(  A_{3}\right)  =\frac{2\lambda-1}{\kappa-2\lambda
}$ and eigenvalues $e_{1,2}\left(  A_{3}\right)  =\frac{3\left(
1-\kappa+4\lambda\right)  }{4\left(  \kappa-2\lambda\right)  }\pm\sqrt
{\frac{9\left(  1-7\kappa^{2}+8\lambda\left(  7+10\lambda\right)
-\kappa\left(  26+24\lambda\right)  \right)  }{16\left(  \kappa-2\lambda
\right)  ^{2}}-\frac{3\left(  1+\kappa\right)  \left(  1+2\lambda\right)
}{\omega}-\frac{27\omega}{2\left(  \kappa-2\lambda\right)  ^{2}}}$. The point
exist for $\kappa\neq2\lambda$ and it describes an asymptotic solution where
the two fluids contribute in the cosmological evolution. The asymptotic
solution describes an acceleration universe for specific values of the free
parameters $\kappa$ and $\lambda$. The stationary point is an attractor when
the real components of the two eigenvalues are negative, that is, when
$\lambda\neq-\frac{1}{2}$, and $\kappa\left(  1+\kappa\right)  +3\omega
<2\lambda\left(  1+\kappa\right)  ~,~\sqrt{\left(  1-\kappa+4\lambda\right)
^{3}\left(  1-49\kappa+100\lambda\right)  }+\kappa\left(  26+7\kappa
+24\lambda\right)  +48\omega>1+8\lambda\left(  7+10\lambda\right)  $ or
$1+\sqrt{\left(  1-\kappa+4\lambda\right)  ^{3}\left(  1-49\kappa
+100\lambda\right)  }+8\lambda\left(  7+10\lambda\right)  <\kappa\left(
26+7\kappa+24\lambda\right)  +48\omega$,~$2\left(  \kappa-2\lambda\right)
\left(  1+2\lambda\right)  +3\omega>0$. In the special case where $\kappa=-1$,
the stationary point is attractor when $\frac{2}{3}<\frac{\omega}{\left(
1+2\lambda\right)  ^{2}}<\frac{5}{6}$.

\item
\[
A_{4}^{\pm}=\left(  \pm\frac{i}{\sqrt{\omega}},0\right)  ,
\]
with $\Omega_{m}\left(  A_{4}^{\pm}\right)  =0$, $w_{eff}^{\Gamma_{1}}\left(
A_{4}^{\pm}\right)  =1\pm\frac{2i}{3\sqrt{\omega}}\sqrt{\frac{2}{3}}$ and
eigenvalues $e_{1}\left(  A_{4}^{\pm}\right)  =3\pm i\left(  1+2\lambda
\right)  \sqrt{\frac{6}{\omega}}$,~$e_{2}\left(  A_{4}^{\pm}\right)  =6\pm
i\left(  1+\kappa\right)  \sqrt{\frac{6}{\omega}}$. The stationary points are
real and physically accepted when $\omega<0$. They describe asymptotic
solutions where only the scalar field contributes in the universe and
acceleration is occurred only at the stationary point $A_{4}^{-}$ for
$-\frac{8}{3}<\omega<0$. Point $A_{4}^{+}$ is an attractor when $\left\{
\left(  1+2\lambda\right)  <-3\sqrt{\frac{\left\vert \omega\right\vert }{6}%
},~\left(  1+\kappa\right)  <-6\sqrt{\frac{\left\vert \omega\right\vert }{6}%
}\right\}  $ while point $A_{4}^{-}$ is an attractor for $\left\{  \left(
1+2\lambda\right)  <3\sqrt{\frac{\left\vert \omega\right\vert }{6}},~\left(
1+\kappa\right)  <6\sqrt{\frac{\left\vert \omega\right\vert }{6}}\right\}  $.
\end{itemize}

The above results are summarized in table \ref{tabl1}.%

\begin{table}[tbp] \centering
\caption{Stationary points and physical parameters for dynamical system $DS_{I}$.}%
\begin{tabular}
[c]{cccccc}\hline\hline
\textbf{Point} & \textbf{Existence} & \textbf{Interaction} & $\mathbf{\Omega
}_{m}$ & $\mathbf{w}_{eff}^{\Gamma_{1}}$ & \textbf{Can be Attractor?}\\\hline
$A_{1}$ & $\omega\neq0$ & Yes & $1+\frac{2\left(  1+2\lambda\right)  ^{2}%
}{3\omega}$ & $\frac{2-8\lambda^{2}}{\omega}$ & Yes\\
$A_{2}$ & $\omega\neq0$ & No & $0$ & $-1+\frac{1-\kappa^{2}}{3\omega}$ & Yes\\
$A_{3}$ & $\kappa\neq2\lambda$ & Yes & $\frac{\left(  1+\kappa\right)  \left(
\kappa-2\lambda\right)  +3\omega}{\left(  \kappa-2\lambda\right)  ^{2}}$ &
$\frac{2\lambda-1}{\kappa-2\lambda}$ & Yes\\
$A_{4}^{\pm}$ & $\omega<0$ & No & $0$ & $1\pm\frac{2i}{3\sqrt{\omega}}%
\sqrt{\frac{2}{3}}$ & Yes\\\hline\hline
\end{tabular}
\label{tabl1}%
\end{table}%

\subsection{Dynamical system $DS_{II}$}

For the second connection, namely $\Gamma_{2}$, the evolution of the physical
variables is given by the dynamical system $DS_{II}$.

The stationary points $B=\left(  x\left(  B\right)  ,y\left(  B\right)
,z\left(  B\right)  \right)  $ for the dynamical system (\ref{ds.04}%
)-(\ref{ds.07}) describe asymptotic solutions with equation of state parameter
for the cosmological fluid $w_{eff}^{\Gamma_{2}}\left(  x,y,z\right)
=y-x\left(  \omega x+z-\frac{2\sqrt{6}}{3}\right)  $.

We calculate the stationary points:

\begin{itemize}
\item
\[
B_{1}=\left(  x_{1},0,-\left(  \frac{1}{x_{1}}+\omega x_{1}\right)  \right)
,
\]
with $\Omega_{m}\left(  B_{1}\right)  =0$ and $w_{eff}^{\Gamma_{2}}\left(
B_{2}\right)  =1+2\sqrt{\frac{2}{3}}x_{1}$. Parameter $x_{1}$ is arbitrary,
which means that $B_{1}$ $\ $describes a family of points. The eigenvalues of
the linearized system are derived $e_{1}\left(  B_{1}\right)  =0$%
,~$e_{2}\left(  B_{1}\right)  =6+\sqrt{6}\left(  1+\kappa\right)  x_{1}~$and
$e_{3}\left(  B_{3}\right)  =3+\sqrt{6}\left(  1+2\lambda\right)  x_{1}$. The
asymptotic solutions given by the family of points $B_{1}$ describe cosmic
acceleration when $x_{1}<-\sqrt{\frac{2}{3}}$. These points have been derived
before in \cite{pal2}, and the application of the Center Manifold Theorem
indicates that there is not any stable submanifold where the points $B_{1}$
are stable.

\item
\[
B_{2}=\left(  0,-1,\sqrt{\frac{2}{3}}\left(  1+\kappa\right)  \right)  ,
\]
where $\Omega_{m}\left(  B_{2}\right)  =0$, $w_{eff}^{\Gamma_{2}}\left(
B_{2}\right)  =-1$, and eigenvalues $e_{1}\left(  B_{2}\right)  =-3$,
$e_{2}\left(  B_{2}\right)  =-3$ and $e_{3}\left(  B_{3}\right)  =-3$. Point
$B_{2}$ describe the de Sitter universe and it is always an attractor.

\item
\[
B_{3}=\left(  0,0,2\sqrt{\frac{2}{3}}\left(  1+2\lambda\right)  \right)  ,
\]
where $\Omega_{m}\left(  B_{3}\right)  =1$, $w_{eff}^{\Gamma_{3}}\left(
B_{2}\right)  =0$, and eigenvalues $e_{1}\left(  B_{3}\right)  =3$%
,~$e_{2}\left(  B_{3}\right)  =-\frac{3}{2}$ and $e_{3}\left(  B_{3}\right)
=-\frac{3}{2}$. The asymptotic solution at point $B_{3}$ describe a universe
dominated by the dust fluid. $B_{3}$ is always a saddle point which means that
the asymptotic solution is unstable.
\end{itemize}

The above results are summarized in table \ref{tab2}.%

\begin{table}[tbp] \centering
\caption{Stationary points and physical parameters for dynamical system $DS_{II}$.}%
\begin{tabular}
[c]{cccccc}\hline\hline
\textbf{Point} & \textbf{Existence} & \textbf{Interaction} & $\mathbf{\Omega
}_{m}$ & $\mathbf{w}_{eff}^{\Gamma_{1}}$ & \textbf{Can be Attractor?}\\\hline
$B_{1}$ & Always & No & $0$ & $1+2\sqrt{\frac{2}{3}}x_{1}$ & No\\
$B_{2}$ & Always & No & $0$ & $-1$ & Yes\\
$B_{3}$ & Always & No & $1$ & $0$ & No\\\hline\hline
\end{tabular}
\label{tab2}%
\end{table}%

\subsection{Dynamical system $DS_{III}$}

For the dynamical system $DS_{III}$ of connection $\Gamma_{3}$, the stationary
points $C=\left(  x\left(  C\right)  ,y\left(  C\right)  ,v\left(  C\right)
\right)  $ describe asymptotic solutions with $w_{eff}^{\Gamma_{3}}\left(
x,y,v\right)  =y+x\left(  \frac{2}{3}\sqrt{6}-\omega x+\frac{1}{3}z\right)  $.

\begin{figure}[ptb]
\centering\includegraphics[width=0.5\textwidth]{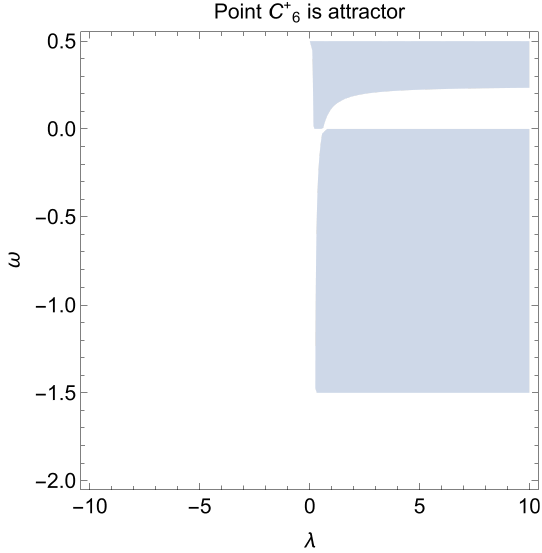}\caption{Region plot
in the two-dimensional plane $\left\{  \lambda,\omega\right\}  $, where point
$C_{6}^{+}$ is attractor.}%
\label{fig01}%
\end{figure}

The stationary points of the dynamical system (\ref{ds.08})-(\ref{ds.11}) are:

\begin{itemize}
\item
\[
C_{1}=\left(  0,-1,-\sqrt{\frac{2}{3}}\left(  1+\kappa\right)  \right)  ,
\]
from where we calculate $\Omega_{m}\left(  C_{1}\right)  =0$, $w_{eff}%
^{\Gamma_{3}}\left(  C_{1}\right)  =-1$ and the three eigenvalues
$e_{1}\left(  C_{1}\right)  =-5$, $e_{2}\left(  C_{1}\right)  =-3$%
,$~e_{3}\left(  C_{1}\right)  =2$. Point $C_{1}$ describes a de Sitter
universe where the scalar field plays the role of the cosmological constant.
From the eigenvalues we infer that the point is always a saddle point.

\item
\[
C_{2}=\left(  0,0,-2\sqrt{\frac{2}{3}}\left(  1+2\lambda\right)  \right)  ,
\]
describes a universe dominated by the dust fluid, $\Omega_{m}\left(
C_{2}\right)  =1$ and $w_{eff}^{\Gamma_{3}}\left(  C_{2}\right)  =0$. The
eigenvalues of the linearized system are $e_{1}\left(  C_{2}\right)  =3$,
$e_{2}\left(  C_{2}\right)  =3$,$~e_{3}\left(  C_{2}\right)  =-\frac{1}{2}$
from where infer that $C_{2}$ is always a saddle point.

\item
\[
C_{3}=\left(  \frac{1+\kappa}{\sqrt{6}\omega},-1-\frac{\left(  1+\kappa
\right)  ^{2}}{6\omega},0\right)  ,
\]
with physical variables $\Omega_{m}\left(  C_{3}\right)  =0$ and
$w_{eff}^{\Gamma_{3}}\left(  C_{3}\right)  =-1+\frac{1-\kappa^{2}}{3\omega}$.
The asymptotic solution describes a universe dominated by the scalar field,
while acceleration can be occurred for specific values of the free parameters.
The eigenvalues of the linearized system are $e_{1}\left(  C_{3}\right)
=-3-\frac{\left(  1+\kappa\right)  \left(  \kappa-2\lambda\right)  }{\omega}$,
$e_{2}\left(  C_{3}\right)  =-3-\frac{\left(  1+\kappa\right)  ^{2}}{2\omega}$
and $e_{3}\left(  C_{3}\right)  =\frac{\kappa\left(  2+3\kappa\right)
-1+10\omega}{6\omega}$. Hence, point $C_{3}$ is an attractor when $\left\{
\kappa<-1,~\frac{\left(  1+\kappa\right)  \left(  3\kappa-1\right)  }%
{10}<\omega<-\frac{\left(  1+\kappa\right)  ^{2}}{6},~\lambda<\frac
{\kappa\left(  1+\kappa\right)  +3\omega}{2\left(  1+\kappa\right)  }\right\}
$ or $\left\{  -1<\kappa<-\frac{1}{3},~0<\omega<\frac{\left(  1+\kappa\right)
\left(  3\kappa-1\right)  }{10}\right\}  $ or $\left\{  \kappa>2,~\frac
{\left(  1+\kappa\right)  \left(  3\kappa-1\right)  }{10}<\omega
<-\frac{\left(  1+\kappa\right)  ^{2}}{6},~\lambda>\frac{\kappa\left(
1+\kappa\right)  +3\omega}{2\left(  1+\kappa\right)  }\right\}  $ .

\item
\[
C_{4}=\left(  \sqrt{\frac{2}{3}}\frac{1+2\lambda}{\omega},0,0\right)  ,
\]
where $\Omega_{m}\left(  C_{4}\right)  =1+\frac{2+8\lambda\left(
1+\lambda\right)  }{3\omega}$, $w_{eff}^{\Gamma_{3}}\left(  C_{4}\right)
=\frac{2\left(  1-4\lambda^{2}\right)  }{3\omega}$ and eigenvalues
$e_{1}\left(  C_{4}\right)  =3+\frac{2\left(  \kappa-2\lambda\right)  \left(
1+2\lambda\right)  }{\omega}$,~$e_{2}\left(  C_{4}\right)  =\frac{1}{6}%
+\frac{4\lambda\left(  1+3\lambda\right)  -1}{3}$,~$e_{3}\left(  C_{4}\right)
=-\frac{3}{2}-\frac{4\lambda\left(  1+\lambda\right)  +1}{\omega}$. This point
describes a solution where the scalar field and the dust fluid contributes in
the cosmological solution. Point $C_{4}$ is an attractor if the free
parameters are constraint as follows $\left\{  \lambda<-\frac{1}{2}%
,\frac{\left(  1-6\lambda\right)  }{1+2\lambda}<\frac{\omega}{\left(
1+2\lambda\right)  ^{2}}<-\frac{2}{3},\kappa<\frac{4\lambda\left(
1+2\lambda\right)  -3\omega}{2\left(  1+2\lambda\right)  }\right\}  $ or
$\left\{  -\frac{1}{2}<\lambda<\frac{1}{6},0<\frac{\omega}{\left(
1+2\lambda\right)  ^{2}}<2\frac{\left(  1-6\lambda\right)  }{\left(
1+2\lambda\right)  },\kappa<\frac{4\lambda\left(  1+2\lambda\right)  -3\omega
}{2\left(  1+2\lambda\right)  }\right\}  $ or $\left\{  \lambda>\frac{1}%
{4},\frac{\left(  1-6\lambda\right)  }{1+2\lambda}<\frac{\omega}{\left(
1+2\lambda\right)  ^{2}}<-\frac{2}{3},\kappa>\frac{4\lambda\left(
1+2\lambda\right)  -3\omega}{2\left(  1+2\lambda\right)  }\right\}  $.

\item
\[
C_{5}^{\pm}=\left(  \pm\frac{i}{\sqrt{\omega}},0,0\right)  ,
\]
exist when $\omega<0$. The physical properties of the asymptotic solutions are
similar to that of points $A_{4}^{\pm}$, that is, $\Omega_{m}\left(
C_{5}^{\pm}\right)  =0$ and $w_{eff}^{\Gamma_{3}}\left(  C_{5}^{\pm}\right)
=1\pm\frac{2i}{3\sqrt{\omega}}\sqrt{\frac{2}{3}}$. We calculate the
eigenvalues of the linearized system $e_{1}\left(  C_{5}^{\pm}\right)  =3\pm
i\left(  1+2\lambda\right)  \sqrt{\frac{6}{\omega}}$,~$e_{2}\left(  C_{5}%
^{\pm}\right)  =6\pm i\left(  1+\kappa\right)  \sqrt{\frac{6}{\omega}}$ and
$e_{3}\left(  C_{5}^{\pm}\right)  =-\left(  \frac{4}{3}\pm2i\sqrt{\frac
{2}{3\omega}}\right)  $. Thus, point $C_{5}^{+}$ is attractor for $\left(
1+2\lambda\right)  <-3\sqrt{\frac{\left\vert \omega\right\vert }{6}},~\left(
1+\kappa\right)  <-6\sqrt{\frac{\left\vert \omega\right\vert }{6}}$, while
point $C_{5}^{-}$ is attractor for~$\left\{  \left(  1+2\lambda\right)
<3\sqrt{\frac{\left\vert \omega\right\vert }{6}},~\left(  1+\kappa\right)
<6\sqrt{\frac{\left\vert \omega\right\vert }{6}},~\omega<-\frac{3}{2}\right\}
$.

\item
\[
C_{6}^{\pm}=\left(  \frac{1\pm\sqrt{1-2\omega}}{\sqrt{6}\omega},0,\sqrt
{\frac{2}{3}}\left(  -2\pm\sqrt{1-2\omega}\right)  \right)  ,
\]
where $\Omega_{m}\left(  C_{6}^{\pm}\right)  =0$ and $w_{eff}^{\Gamma_{3}%
}\left(  C_{6}^{\pm}\right)  =\frac{1}{9}+\frac{2\left(  1\pm\sqrt{1-2\omega
}\right)  }{\omega}$. The points are real when $\omega\leq\frac{1}{2}$, where
easy we can infer that only the asymptotic solution at point $C_{6}^{+}$ can
describe cosmic acceleration for $-\frac{3}{2}<\omega<0$. The eigenvalues are
studied numerically. We find that point $C_{6}^{-}$ can not be an attractor,
while point $C_{6}^{+}$ is attractor when $\left\{  \omega<0,~\kappa>-\frac
{4}{3}+\frac{5}{3}\sqrt{1-2\omega}\right\}  $ or $\left\{  0<\omega<\frac
{1}{2},~\kappa<-\frac{4}{3}+\frac{5}{3}\sqrt{1-2\omega}\right\}  $ and
$\lambda$ is constraint as given in Fig. \ref{fig01}.

\item
\[
C_{7}=\left(  \frac{5}{1-3\kappa}\sqrt{\frac{2}{3}},\frac{2\left(
9\kappa\left(  8+3\kappa\right)  +50\omega-9\right)  }{3\left(  1-3\kappa
\right)  ^{2}},\frac{\sqrt{6}\left(  1-\kappa\left(  2+3\kappa\right)
+10\omega\right)  }{2\left(  1-3\kappa\right)  }\right)  ,
\]
with $\Omega_{m}\left(  C_{7}\right)  =0$ and $w_{eff}^{\Gamma_{3}}\left(
C_{7}\right)  =\frac{\kappa-7}{3\left(  3\kappa-1\right)  }$. The point exist
for $\kappa\neq\frac{1}{3}$ and describes an accelerated universe for
$\frac{1}{3}<\kappa<2$. The eigenvalues are determined numerically. In Fig.
\ref{fig.02} we present region plots in the space of the free parameters
$\left\{  \kappa,\lambda,\omega\right\}  $ where the stationary point is attractor.

\item
\[
C_{8}=\left(  -\frac{1}{\kappa-2\lambda}\sqrt{\frac{2}{3}},\frac{2\left(
\kappa-2\lambda\right)  \left(  1+2\lambda\right)  +3\omega}{2\left(
\kappa-2\lambda\right)  ^{2}},0\right)  ,
\]
from where we calculate $\Omega_{m}\left(  C_{8}\right)  =\frac{\left(
1+\kappa\right)  \left(  \kappa-2\lambda\right)  +3\omega}{\left(
\kappa-2\lambda\right)  }$ and $w_{eff}^{\Gamma_{3}}\left(  C_{8}\right)
=\frac{2\lambda-1}{\kappa-2\lambda}$. The stationary point exist for
$\kappa-2\lambda\neq0$ and it can describe cosmic acceleration for specific
values of the free parameters $\kappa,\lambda$. \ The eigenvalues are studied
numerically. In Fig. \ref{fig.03} we present region plots in the space of the
free parameters $\left\{  \kappa,\lambda,\omega\right\}  $ where the
stationary point $C_{8}$ is attractor.

\item
\[
C_{9}=\left(  \frac{1}{\sqrt{6}\left(  1-6\lambda\right)  },0,\frac{\sqrt
{6}\left(  \omega-2+8\lambda\left(  1+3\lambda\right)  \right)  }{2\left(
1-6\lambda\right)  }\right)  ,
\]
with $\Omega_{m}\left(  C_{9}\right)  =\frac{2\left(  12\lambda\left(
6\lambda-1\right)  +\omega\right)  }{3\left(  1-6\lambda\right)  ^{2}}$ and
$w_{eff}^{\Gamma_{3}}\left(  C_{9}\right)  =\frac{1-2\lambda}{3\left(
1-6\lambda\right)  }$. The stationary point exist for $\lambda\neq\frac{1}{6}$
and it describes an accelerated universe for $\frac{1}{6}<\lambda<\frac{1}{4}%
$. The eigenvalues are studied numerically. In Fig. \ref{fig.04} we present
region plots in the space of the free parameters $\left\{  \kappa
,\lambda,\omega\right\}  $ where the stationary point $C_{9}$ is attractor.
\end{itemize}

The above results are summarized in table \ref{tab3}.

\begin{figure}[ptb]
\centering\includegraphics[width=1\textwidth]{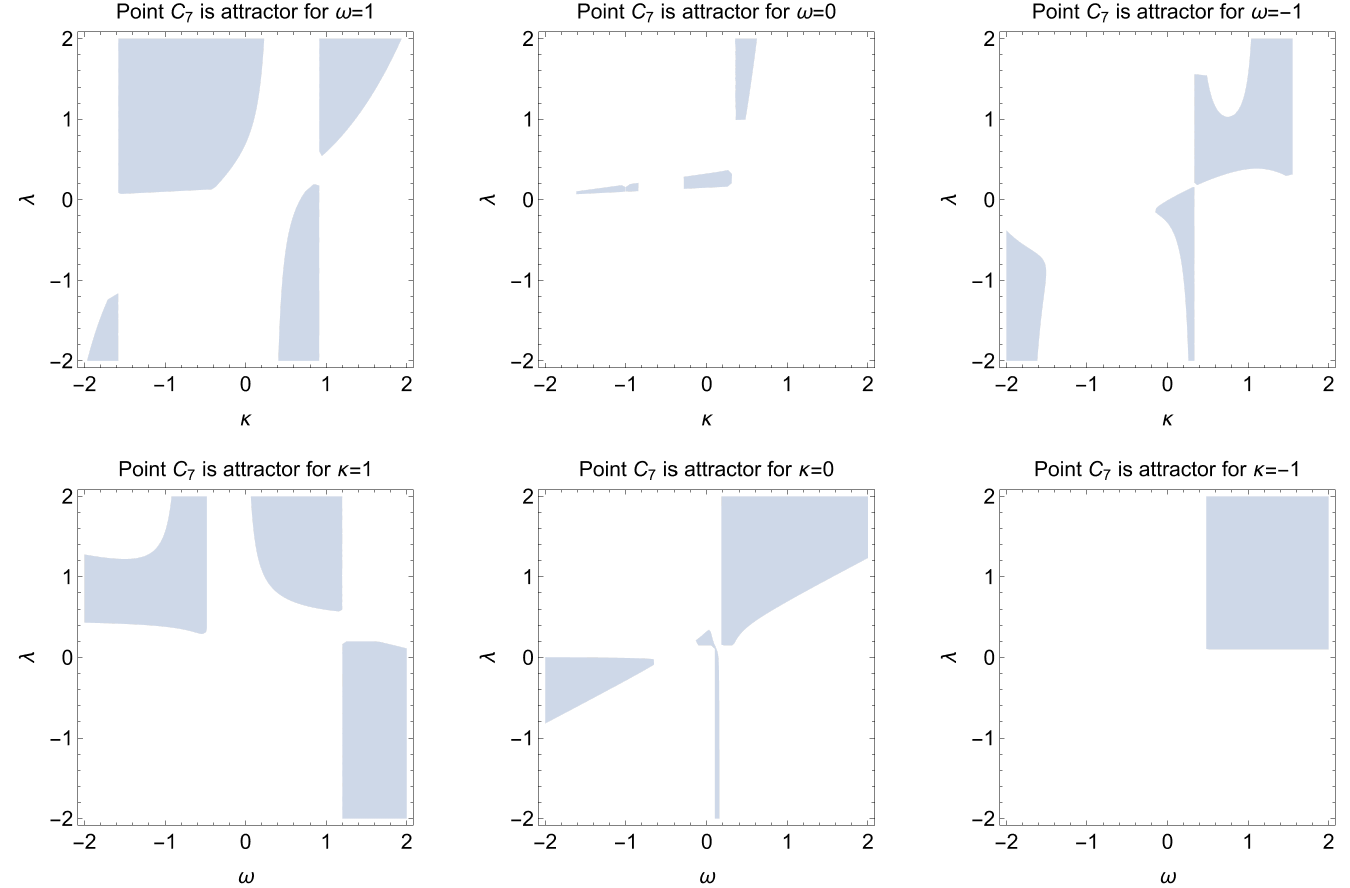}\caption{Region plots
in the space of the free parameters $\left\{  \kappa,\lambda,\omega\right\}  $
where the three eigenvalues of the dynamical system $DS_{III}$ around the
stationary point $C_{7}$ have negative real parts and $C_{7}$ is attractor.}%
\label{fig.02}%
\end{figure}

\begin{figure}[ptb]
\centering\includegraphics[width=1\textwidth]{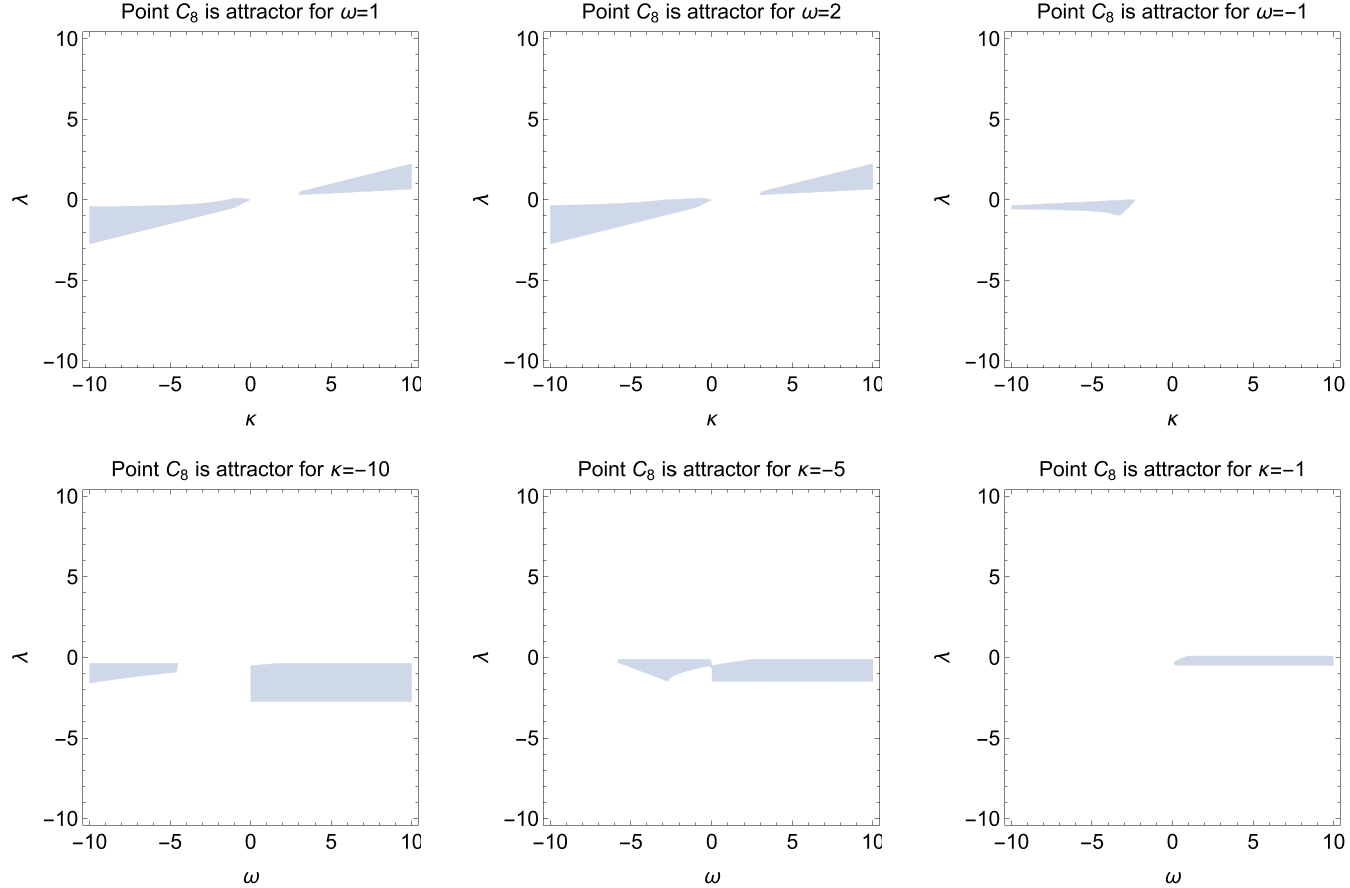}\caption{Region plots
in the space of the free parameters $\left\{  \kappa,\lambda,\omega\right\}  $
where the three eigenvalues of the dynamical system $DS_{III}$ around the
stationary point $C_{8}$ have negative real parts and $C_{8}$ is attractor.}%
\label{fig.03}%
\end{figure}

\begin{figure}[ptb]
\centering\includegraphics[width=1\textwidth]{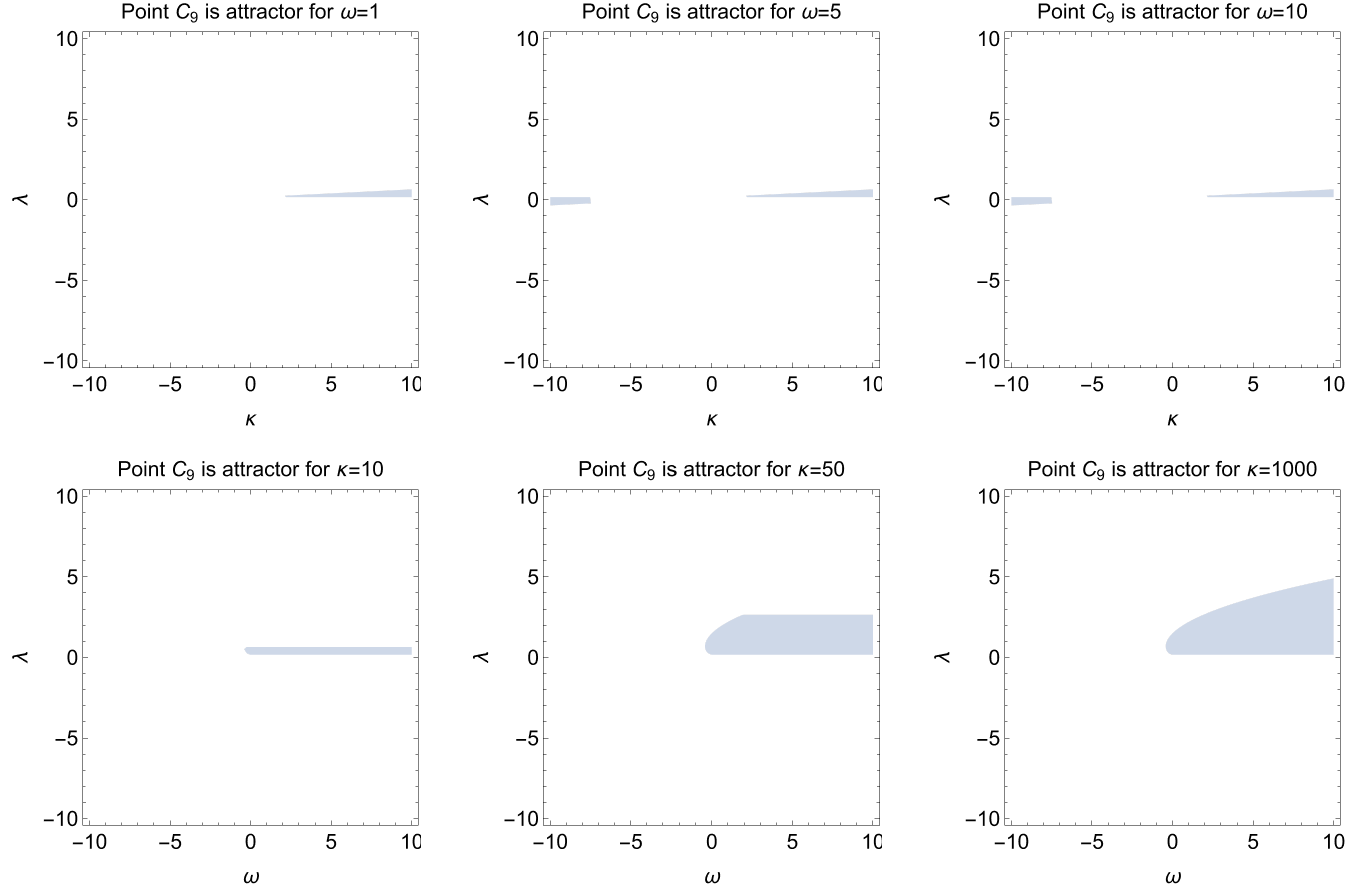}\caption{Region plots
in the space of the free parameters $\left\{  \kappa,\lambda,\omega\right\}  $
where the three eigenvalues of the dynamical system $DS_{III}$ around the
stationary point $C_{8}$ have negative real parts and $C_{8}$ is attractor.}%
\label{fig.04}%
\end{figure}%

\begin{table}[tbp] \centering
\caption{Stationary points and physical parameters for dynamical system $DS_{III}$.}%
\begin{tabular}
[c]{cccccc}\hline\hline
\textbf{Point} & \textbf{Existence} & \textbf{Interaction} & $\mathbf{\Omega
}_{m}$ & $\mathbf{w}_{eff}^{\Gamma_{1}}$ & \textbf{Can be Attractor?}\\\hline
$C_{1}$ & Always & No & $0$ & $-1$ & No\\
$C_{2}$ & Always & No & $1$ & $0$ & No\\
$C_{3}$ & $\omega\neq0$ & No & $0$ & $-1+\frac{1-\kappa^{2}}{3\omega}$ & Yes\\
$C_{4}$ & $\omega\neq0$ & Yes & $1+\frac{2+8\lambda\left(  1+\lambda\right)
}{3\omega}$ & $\frac{2\left(  1-4\lambda^{2}\right)  }{3\omega}$ & Yes\\
$C_{5}$ & $\omega<0$ & No & $0$ & $1\pm\frac{2i}{3\sqrt{\omega}}\sqrt{\frac
{2}{3}}$ & Yes\\
$C_{6}^{\pm}$ & $\omega\leq\frac{1}{2}$ & No & $0$ & $\frac{1}{9}%
+\frac{2\left(  1\pm\sqrt{1-2\omega}\right)  }{\omega}$ & $C_{6}^{+}$ Yes\\
$C_{7}$ & $\kappa\neq\frac{1}{3}$ & No & $0$ & $\frac{\kappa-7}{3\left(
3\kappa-1\right)  }$ & Yes\\
$C_{8}$ & $\kappa\neq2\lambda$ & Yes & $\frac{\left(  1+\kappa\right)  \left(
\kappa-2\lambda\right)  +3\omega}{\left(  \kappa-2\lambda\right)  }$ &
$\frac{2\lambda-1}{\kappa-2\lambda}$ & Yes\\
$C_{9}$ & $\lambda\neq\frac{1}{6}$ & Yes & $\frac{2\left(  12\lambda\left(
6\lambda-1\right)  +\omega\right)  }{3\left(  1-6\lambda\right)  ^{2}}$ &
$\frac{1-2\lambda}{3\left(  1-6\lambda\right)  }$ & Yes\\\hline\hline
\end{tabular}
\label{tab3}%
\end{table}%

\section{Arbitrary potential}

\label{sec7a}

Let us now focus in the case of the exponential coupling function (\ref{cp1})
with the arbitrary scalar field potential $V\left(  \phi\right)  $ where now
parameter $\kappa$ is not a constant. Specifically, the evolution of parameter
$\kappa$ is given by the first-order differential equation%
\begin{equation}
\frac{d\kappa}{d\tau}=\sqrt{6}x\kappa^{2}\left(  \Lambda\left(  \kappa\right)
-1\right)  ~,~\Lambda\left(  \kappa\right)  =\frac{V_{,\phi\phi}V}{\left(
V_{,\phi}\right)  ^{2}}. \label{nw1}%
\end{equation}
Hence, for an arbitrary potential $V\left(  \phi\right)  $, the dimension of
the dynamical systems $DS_{I}$,~$DS_{II}$ and $DS_{III}$ is increased by one,
where the new equation is (\ref{nw1}). The stationary points of the dynamical
systems $DS_{I}$,~$DS_{II}$ and $DS_{III}$ should satisfy the additional
constraint $\sqrt{6}x\kappa^{2}\left(  \Lambda\left(  \kappa\right)
-1\right)  =0$. Thus, if $\kappa_{0}$ is a root of the algebraic equation
$\kappa_{0}^{2}\left(  \Lambda\left(  \kappa_{0}\right)  -1\right)  =0$, then
the previous stationary points for the exponential potential are recovered.
Additional, the new stationary point which is introduced in the dynamical
system is that with $x=0$. For the dynamical system $DS_{I}$ stationary points
with $x=0$, exist only when~$\kappa=-1$ or $\lambda=-\frac{1}{2}$. For the
dynamical system $DS_{II}$, we find only the stationary points $B_{2}$ and
$B_{3}$, while for the dynamical system $DS_{III}$ the stationary points
$C_{1}$ and $C_{2}$ exists.\ Hence, the physical properties of the asymptotic
solutions remain the same with the exponential potential. However, because of
the new dynamical variable, the stability properties are affected by the
definition of function $\Lambda\left(  \kappa\right)  $.

If we consider the general coupling function $f\left(  \phi\right)  $, then,
$\lambda$ is a dynamical variables and satisfies the equation
\begin{equation}
\frac{d\lambda}{d\tau}=\sqrt{6}x\lambda^{2}\left(  \hat{\Lambda}\left(
\lambda\right)  -1\right)  ~,~\hat{\Lambda}\left(  \lambda\right)
=\frac{f_{,\phi\phi}f}{\left(  f_{,\phi}\right)  ^{2}}\text{.} \label{nw2}%
\end{equation}
When the scalar field potential $V\left(  \phi\right)  $ is the exponential
function, such that $\kappa$ is constant, then the stationary points for the
extended dynamical systems $DS_{I}$,~$DS_{II}$ and $DS_{III}$ are similar with
that which discussed before. Nevertheless, when $\kappa~$and $\lambda$ are not
constants, parameters $\lambda$ and $\kappa$ are not independent, and we can
express $\lambda=\lambda\left(  \kappa\right)  $. However, the families of the
stationary points are these which we have already discussed. Hence, our main
conclusions holds and for other functional forms for the scalar field
potential and the coupling function.

\section{Conclusions}

\label{sec7}

In this study, we considered an interaction within the components of the dark
sector of the universe within the framework of gravitational theory with
nonmetricity. Specifically, we employed a pressureless fluid source to
describe the dark matter component of the universe, coupled with a scalar
field that is nonminimally coupled to the nonmetricity tensor to account for
the universe's acceleration. To model this interaction, we introduced a
coupling function in the Action Integral, which gives rise to a Chameleon mechanism.

For this gravitational model, we conducted a comprehensive analysis of the
dynamics governing the cosmological field equations within a spatially flat
FLRW geometry. We found that the cosmological history depends on the selection
of the connection which defines the nonmetricity scalar. We remark that the
definition of a flat and symmetric connection in FLRW geometry is not unique,
and for the spatially flat geometry there are three families of connections.

Connection $\Gamma_{1}$ provides the cosmological model with the fewest number
of dynamical variables. Utilizing the Hubble normalization approach, we
computed the stationary points for this connection. The field equations
corresponding to connection $\Gamma_{1}$ yield four families of stationary
points as summarized in Table \ref{tabl1}. Two of these families describe
universes with a nontrivial interaction term, while the rest two points when
they exist describe scaling solutions, which can correspond to the asymptotic
behaviour of the early or time acceleration phases of the universe.

Moving to the second connection, $\Gamma_{2}$, we found that the field
equations admit three families of stationary points as presented in Table
\ref{tab2}. However, none of these points describe asymptotic solutions with
an interaction between the matter components of the universe. Lastly, the
field equations of connection $\Gamma_{3}$ allow for the maximum number of
stationary points, with nine families of such points identified with physical
properties are given in Table \ref{tab3}. There are three stationary points
with nonzero interaction term, one stationary point where the matter source
dominates in the cosmic evolution and five stationary points where the scalar
field dominates in the universe.

Among these, three families support the Chameleon mechanism. It is important
to note that our analysis was performed within a finite regime. Although the
dynamical variables are not constrained and can take values at infinity, we
omitted the analysis at infinity due to our finding that the stationary points
of the vacuum scalar nonmetricity model are recovered in this regime.

The findings from this study can serve as selection rules for the connection
in nonmetricity gravity when interaction is required in the gravitational
model. For connection $\Gamma_{2}$ we did not find any stationary point with
nonzero contribution of the Chameleon mechanism. From the Klein-Gordon
equation (\ref{cn.07}) of the second scalar field, it follows that the scalar
field has an exact solution, and there is not any contribution of the matter
source to the mass of the scalar field. It is clear that either in the
perturbation level that will not change. On the other hand, for the rest two
connections, namely, connections while for connections $\Gamma_{1}$ and
$\Gamma_{3}$, the admitted stationary points can describe the major eras of
the cosmological evolution, while they provide asymptotic solutions with
interacting fluids. 

Indeed, coexistence and nonzero interaction in the dark sector of the universe
is provided only by the connections $\Gamma_{1}$ and $\Gamma_{3}$. Thus, this
study can be used to define criteria for the selection of the connection in
nonmetricity theory. Indeed, if interaction is essential in the dark sector,
then connection $\Gamma_{2}$ is excluded from being physically accepted. 

At this stage it is important to mention that because of the introduction of
the Chameleon mechanism, that is, of the nonzero parameter $\lambda$, new
stationary points of the cosmological field equations exist, with different
physical properties and stability behaviour for the majority of the stationary
points, either for that which does not describe coexistence and they do not
support a Chameleon term. 

Recent studies \cite{per1,per2} on the analysis of the cosmological
perturbations in $f\left(  Q\right)  $-gravity indicate the pathological
character of the theory because of the presence of ghost or of the strong
coupling problem. Although the symmetric teleparallel Brans-Dicke model
(\ref{ai.002a}) is equivalent to $f\left(  Q\right)  $-gravity when $\omega
=0$, in the cosmological perturbation theory, for $\omega\neq0$, the scalar
field perturbations provides dynamical degrees of freedom, which may overpass
the pathologies of $f\left(  Q\right)  $-theory. On the other hand, the
existence of the Chameleon mechanism can affect the dynamical evolution \ for
the cosmological perturbations \cite{mot1}, such that to solve the strong
coupling problem. However, such analysis is out of the scopus of this work and
will be studied elsewhere.

\begin{acknowledgments}
AP thanks the support of VRIDT through Resoluci\'{o}n VRIDT No. 096/2022 and
Resoluci\'{o}n VRIDT No. 098/2022.
\end{acknowledgments}

\end{document}